\documentstyle[12pt,epsf]{article}
\newlength{\dinwidth} 
\newlength{\dinmargin}
\begin{document}
%
\def\g{\gamma}
\def\xg{x_\g}
\def\xgo{x_\g^{{\rm obs}}}
\def\xgom{x_\g^{{\rm cal}}}
\def\xglo{x_\g^{{\rm LO}}}
\def\xp{x_p}
\def\xpo{x_p^{{\rm obs}}}
\def\xplo{x_p^{{\rm LO}}}
\def\pT{p_T}
\def\pZ{p_{\it z}}
\def\Z{{\it z}}
\def\ra{\rightarrow}
\def\ETAJ{\eta^{{\rm jet}}}
\def\ETJ{E_T^{{\rm jet}}}
\def\ETAP{\eta^{{\rm parton}}}
\def\ETP{E_T^{{\rm parton}}}
\def\ETJM{E_T^{{\rm cal}}}
\def\DETA{|\Delta\eta|}
\def\ETAM{\eta^{{\rm cal}}}
\def\PHIM{\phi^{{\rm cal}}}
\def\EEP{E^\prime_{e}}
\def\TEP{\theta^\prime_{e}}
\newcommand{\gsim}{\buildrel{>}\over{\sim}}
\def\lapproxeq{\lower .7ex\hbox{$\;\stackrel{\textstyle <}{\sim}\;$}}
\def\gapproxeq{\lower .7ex\hbox{$\;\stackrel{\textstyle >}{\sim}\;$}}
\def\sgp{s_{\gamma p}}
\def\ptmin{p_T^{{\rm min}}}
\def\PRES{P_{{\rm res}}}

\begin{titlepage}

\title{Multiparton Interactions in Photoproduction at HERA}

\author{J.~M.~Butterworth$^1$, J.~R.~Forshaw$^2$ and M.~H.~Seymour$^3$\\
$^1$\footnotesize University College London, Gower St., London WC1E 6BT, UK. \\
$^2$\footnotesize University of Manchester, Oxford Rd., Manchester, M13 9PL, 
UK.\\ 
$^3$\footnotesize TH Division, CERN, CH-1211 Gen\`eve 23, Switzerland.\\
}

\date{29$^{th}$ Jan. 1996}
\maketitle

\vspace{-10.5 cm}
\begin{flushright}
CERN-TH/95-82\\
MC-TH-96/05\\
UCL/HEP 96-02\\
\end{flushright}

\vspace{10 cm}

\begin{abstract}   
The high energy photoproduction of jets is being observed at the $ep$ collider,
HERA\@. It may be that the HERA centre-of-mass energy is sufficiently large
that the production of more than one pair of jets per $ep$
collision becomes possible, owing to the large number density of the probed 
gluons.
We construct a Monte Carlo model of such multiparton interactions and study
their effects on a wide range of physical observables.
The conclusion is that multiple interactions could have very significant 
effects upon the photoproduction final state and that this would for example 
make extractions of the gluon density in the photon rather difficult. Total
rates for the production of many (i.e.\ $\ge 3$) jets could provide 
direct evidence for the presence of multiple interactions, although parton 
showering and hadronization significantly affect low transverse energy jets.
\end{abstract} 

\setcounter{page}{0}
\thispagestyle{empty}
\newpage

\end{titlepage}

\newpage

\section{Introduction} 
In recent years, the TEVATRON ($\bar{p}p$) and HERA ($ep$)  colliders have 
made it possible to study the standard theory of strong interactions (QCD) 
in a new regime: the regime of high parton densities. QCD predicts 
a rapid build up of very slow gluons within hadrons, which can be observed 
at high centre of mass (CM) energies, i.e.\ $x \sim Q^2/s \ll 1$ where 
$s$ is the CM energy and $Q^2$ is the appropriate hard scale. Both the 
TEVATRON and HERA can study this region in some detail, through the production 
of high $p_T$ jets, heavy flavours and large-$t$ diffractive scattering. 
At HERA, we have the additional possibility of studying 
this `small-$x$' physics using deep inelastic scattering 
(where $x = x_{{\rm Bj}}$).

In this paper, we study jet production at HERA in those events where
the photon is nearly real (photoproduction). Photoproduction of jets has been
observed at HERA~\cite{ZH1} and different components of the cross section 
identified~\cite{ZEUS2}. The components are easy to define in leading order 
QCD and are termed {\it direct\/} and {\it resolved}. In the direct 
sector, the photon carries all of its energy into the hard scatter 
whilst in the resolved
sector, only a fraction of the photon energy participates in the hard 
subprocess. At higher orders this simple distinction between direct and 
resolved is no longer uniquely defined. However, a precise definition of the 
separation between resolved and direct processes in terms of physical
observables is possible~\cite{DIJETS} allowing the effects of the parton 
distributions in the proton and photon to be disentangled \cite{JEFF1}. 
This has already been implemented in higher order dijet cross section
calculations~\cite{NLO}.

The total rate is high for the photoproduction of jets and allows a study of 
small $x$ phenomena when $x \sim {\ETJ}^2/s \ll 1$ ($\ETJ$ is the jet 
transverse momentum). Naively, one might conclude that the 1800~GeV CM energy
available at the TEVATRON would make it far superior to HERA 
(with a $\gamma p$ CM energy that is typically $\simeq 200$ GeV) regarding 
the study of small $x$ physics through jet production. However, this is not 
the case. Associated with higher CM energies is an increase in the background 
of soft physics. At HERA, the asymmetric configuration of the lab frame 
(27 GeV leptons collide with 820~GeV protons) tends to boost the poorly 
understood `soft' physics into the proton direction and hence down the 
`forward' beam hole. The asymmetric boost of the $\gamma p$ system in the 
lab also means that jets that have been produced by small $x$ partons within 
the proton appear in the central region in the lab frame of reference 
and are thus clearly visible in the detectors. None of these benefits are 
present at the TEVATRON, where the lab frame is also the $p\bar{p}$ CM frame.

Due to the proliferation of low $x$ partons (which can be inferred from the
strong growth observed in the HERA data on the proton structure function
\cite{DIS}), it is possible that more than
one pair of jets can be produced per $\g p$ collision. This multiple (parton)
scattering, illustrated in figure~1, is expected at high enough CM energies
and signals the onset of unitarization corrections to the simple perturbative
QCD
picture of $2 \to 2$ parton scattering. Unitarization corrections
are certainly necessary since the total cross section for inclusive jet
photoproduction (calculated in lowest order QCD with steeply rising parton
distribution functions) will ultimately exceed the total $\gamma p$ cross 
section\footnote{Multiple scattering is only one manifestation of the 
unitarization corrections that are expected to occur at high energies. 
The steepness of the parton densities will eventually be tamed by additional 
corrections. We ignore such effects in this paper.}.
As we shall see, this apparent anomaly is
resolved once we appreciate that the inclusive jet cross section exceeds the
total cross section for jet production by a factor equal to the mean
multiplicity of multiple interactions. We introduce an eikonal model which
assumes that individual hard scatters are uncorrelated. This allows us
to model the rate of multiple interactions and study their effects upon
the hadronic final state. For the latter study, it is most convenient to
subject our model to a  Monte Carlo simulation and this will allow us to
make full use of the available and forthcoming HERA data.

\begin{figure}
\begin{center}
\leavevmode
\hbox{\epsfysize=1.75 in
\epsfxsize= 2.5 in
\epsfbox{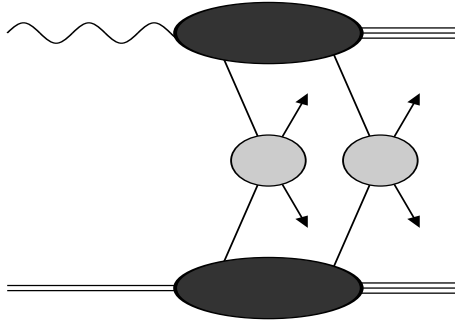}}
\end{center}
\vspace{-5mm}
\caption{An example of a multiple scattering in a $\gamma p$ collision.}
\end{figure}

In section~\ref{sec:model}, we describe the eikonal model of 
refs.\cite{CL,FS,FH} and show how it predicts a significant rate for the 
production of multijet events at HERA energies. Also described here is the 
implementation of the formalism within the HERWIG Monte Carlo package
\cite{HRW}. By integrating our formalism within HERWIG, we can make
realistic studies of the final state which include the effects of parton 
showering and hadronization.  In section~\ref{sec:sm} a number of key 
photoproduction distributions are presented for our default model, and 
compared to the results obtained without multiple interactions. 
In sections 4 and 5, we examine the effects of using different parton
distribution functions and of variations on the default model.
In section 6, comparisons are made to available HERA data. We show that 
multiple scattering can be expected to make a clean extraction of parton 
distribution functions in the photon rather difficult. 

\section{Why Multiple Interactions at HERA?}
\label{sec:model}

We are interested in jet production in $\g p$ reactions and at the level of
the hard subprocess we assume that this can be approximated by the lowest
order matrix element for $2 \to 2$ parton scattering with final state
partons produced with a transverse momentum, 
$\pT > \ptmin \gg \Lambda_{{\rm QCD}}$. Jets can be produced {\it directly\/}
via the $\g$-parton hard subprocess or indirectly via partons from
the {\it resolved\/} photon scattering with partons in the proton. These
resolved partons can be generated either non-perturbatively (i.e.\ the $\g$
splits into a large size $q \bar{q}$ pair) or perturbatively (i.e.\ via
perturbative evolution of a small size $q \bar{q}$ pair).
Let us start by considering a $\g p$ interaction at some fixed
centre-of-mass energy, $\sgp$. In the CM frame
we think of the proton and resolved photon as Lorentz contracted `parton
pancakes' colliding at some impact parameter, $b$.
The mean number of jet pairs produced in this resolved-$\gamma$--$p$ 
interaction is then 
\begin{equation}
\langle n(b,s) \rangle = {\cal L}_{{\rm partons}} \otimes \hat{\sigma}_H
\end{equation}
where ${\cal L}_{{\rm partons}}$ is the parton luminosity and $ \hat{\sigma}_H$
is the cross section for a pair of partons to produce a pair of jets (i.e.\
partons with $p_T > \ptmin$). The direct photon interaction generates only
a single pair of partons and we ignore it for now (it will be included as
a separate hard subprocess in the Monte Carlo simulation).

The convolution is because the parton cross section depends upon the parton
energies. More specifically,
\begin{equation}
d {\cal L}_{{\rm partons}} = A(b) 
n_{\gamma}(x_{\gamma}) n_p(x_p) dx_{\gamma} dx_p
\end{equation}
where $n_i(x_i)$ is the number density of partons in hadron $i$ which carry
a fraction $x_i$ of the hadron energy. For ease of notation we do not
distinguish between parton types and have ignored any scale dependence of
the number densities. $A(b)$ is a function which specifies the
distribution of partons in impact parameter. It must satisfy
$$ \int \pi db^2 A(b) = 1$$ in order that the parton luminosity integrated
over all space is simply the product of the parton number densities.
Factorizing the $b$ dependence like this is an assumption. In particular
we do not contemplate QCD effects which would spoil this, e.g.\ perhaps
leading to \lq hot spots' of partons.
For the proton, the number density is none other than
the proton parton density, i.e.\ $n_p(x_p) \equiv f_p(x_p)$.
The number density $n_{\gamma}$ of partons given that the photon is resolved is
related to the photon parton density by a factor of $\sim  \alpha_{{\rm em}}$,
i.e.\ $n_{\gamma}(x_{\gamma}) \sim f_{\gamma}(x_{\gamma})/\alpha_{{\rm em}}$.
In our model, we ignore multiple scattering which arises due to interactions
of the small size fluctuations of the incoming photon
and therefore assume that the photon interacts just like a hadron. The
large size hadronic fluctuations of the photon can be modelled assuming
vector meson dominance and allow us to estimate that
$n_{\gamma}(x_{\gamma}) = f_{\gamma}(x_{\gamma})/\PRES$
where $\PRES = \kappa 4 \pi \alpha_{{\rm em}}/f_{\rho}^2$. The sum over
hadronic fluctuations determines $\kappa$. Unless otherwise stated, we
assume $\rho$-dominance, i.e.\ $\kappa = 1$ and $\PRES \approx 1/300$.   

Thus, after performing the convolution, we can write:
\begin{equation}
\langle n(b,\sgp) \rangle = \frac{A(b)}{\PRES} \sigma_H^{{\rm inc}}(\sgp),
\end{equation}
where $\sigma_H^{{\rm inc}}(\sgp)$ is the {\it inclusive\/} cross section for
$\g p \to$ jets. Restoring the parton indices, it is given by
\begin{equation}
\sigma^{{\rm inc}}_H(\sgp) = \int_{p_T^{{\rm min}2}}^{\sgp/4}
\hspace{-0.4cm} dp_T^2 \int_{4p_T^2 /\sgp}^{1} \hspace{-0.4cm}
 dx_\gamma \int_{4p_T^2 / x_{\gamma} \sgp}^{1} \hspace{-0.5cm} dx_p \sum_{ij}
f_{i/\gamma }(x_\gamma,p_T^2) f_{j/p}(x_p,p_T^2) \;
\frac{d\hat{\sigma}_{ij}(x_\gamma x_p \sgp,p_T)}{dp_T^2}.
\end{equation}

In order to investigate further the structure of events containing 
multiple interactions we need to know the probability distribution for
having $m$ (and only $m$) scatters in a given resolved-$\gamma$--$p$ event,
$P_m$. In order to do this we assume that the separate scatters are
uncorrelated, i.e.\ they obey Poissonian statistics. 
Thus
\begin{equation}
P_m = \frac{(\langle n(b,\sgp) \rangle)^m}{m!} \exp (-\langle n(b,\sgp)
\rangle).
\end{equation}
This formula is central to the Monte Carlo implementation in {\small HERWIG}.

We can now ask for the total cross section for $\gamma p \to$ partons with
$p_T > \ptmin$.
\begin{eqnarray}
\sigma_H(\sgp) &=& \pi \PRES \int db^2 \sum_{m=1}^{\infty} P_m \nonumber \\
&=& \pi \PRES \int db^2 [1 - \exp(-\langle n(b,\sgp)\rangle)].
\end{eqnarray}

Since the total inclusive cross section ($\sigma_H^{{\rm inc}}$) counts 
all jet pairs (even ones which occur in the same event) we expect it to be 
larger than $\sigma_H$ by a factor equal to the 
mean number of multiple interactions
per event (i.e.\ averaged over impact parameter). This is easy to see. Let
$\langle n(\sgp) \rangle$ be the average number of jet pairs produced in
resolved-$\gamma$--$p$ events which contain at least one pair of jets, then
\begin{eqnarray}
\langle n(\sgp) \rangle &=& \frac{ \int db^2 \sum_{m=1}^{\infty} m P_m }{
\int db^2 \sum_{m=1}^{\infty} P_m} \nonumber \\
&=& \frac{ \int db^2 \langle n(b,\sgp) \rangle }{
\int db^2 [1 - \exp( -\langle n(b,\sgp)\rangle)]} \nonumber \\
&=& \frac{ \sigma_H^{{\rm inc}}(\sgp)}{\sigma_H(\sgp)}.
\end{eqnarray}
Note that $\sigma_H$ must always be less than the total $\gamma p$ cross
section, whereas $\sigma_H^{{\rm inc}}$ need not be.

Jet cross sections at HERA are sensitive to the proton parton density
down to small momentum fractions, $x \sim 10^{-3}$.  As shown by the
measurement of the structure function in deep inelastic scattering at
HERA, this is the region of high parton density (which rises as
$x^{-0.3}$ at $Q^2\sim 10$~GeV$^2$~\cite{DIS}).  For $\ptmin\sim1-2$~GeV,
one expects that $\sigma_H^{{\rm inc}}(\mbox{1 TeV}) \sim 1$~mb,
which is much larger than the {\it total} $\gamma p$ cross section, 
which is $\sim 200\;\mu$b according to
the successful \cite{SIGTOT} Regge pole model of Donnachie and
Landshoff~\cite{DL}.  This suggests the possibility that 
$\langle n(\sgp) \rangle$ is significantly greater than 1.

To complete the description of our model, we must now specify how the
partons are distributed in impact parameter space. As a first approximation,
we ignore pQCD effects and assume that $A(b)$ can be written as a
convolution of form factor distributions, i.e.\ starting from
\begin{eqnarray}
G_p({\bf b}) &=& \int \frac{d^2 {\bf k}}{(2 \pi)^2} 
\frac{\exp({\bf k}\cdot{\bf  b})}{(1+{\bf k}^2/\mu^2)^2} \nonumber \\
G_{\gamma}({\bf b}) &=& \int \frac{d^2 {\bf k}}{(2 \pi)^2}
\frac{\exp({\bf k}\cdot{\bf  b})}{1+{\bf k}^2/\nu^2}
\end{eqnarray}
with $\mu^2 = 0.71$ GeV$^2$ and $\nu^2 = 0.47$ GeV$^2$, we
can write
\begin{equation}
A(b) = \int d^2 {\bf b}' \; G_p({\bf b}') G_{\gamma}({\bf b} - {\bf b}').
\end{equation}
The integral then yields \cite{FH}:
\begin{equation}
A(b) = \frac{1}{2\pi} \frac{\mu^2 \nu^2}{\mu^2-\nu^2} \left[
\frac{\mu^2}{\mu^2-\nu^2}(K_0(\nu b) - K_0(\mu b)) - \frac{\mu b}{2} K_1(\mu b)
\right].
\end{equation}

The formalism just described has been discussed extensively in the literature 
(see ref.\cite{EIK,others}) but largely in the 
context of total cross sections. The expected rates for
multi-jet processes have also been studied~\cite{HONJO}. Owing to the
complexity of the final state, in which many jet pairs may be produced,
previous studies have been confined to rather inclusive properties of 
the model. To really test for the existence of multiple interactions however, 
it is necessary to confront the vast amount of data that HERA is making 
available with more detailed properties of the final state and to this end 
we have constructed a Monte Carlo simulation program. In ref.\cite{JIMMY} 
some results of an earlier version of this simulation were presented. In 
this paper we present a more comprehensive study of the effects of this model 
on the photoproduction final state. The program is written as an extension 
to the HERWIG~\cite{HRW} package and can also be used to simulate 
the effects of multiple interaction in other processes, such as
$\gamma\gamma$ interactions in $e^+ e^-$ collisions \cite{Rod}. There is a 
degree of overlap with the multiple interaction formalism that is 
included in the PYTHIA program~\cite{PYTHIA}. 
 
\begin{figure}
\setlength{\unitlength}{1mm}
\epsfysize=400pt
\epsfbox[50 150 450 550]{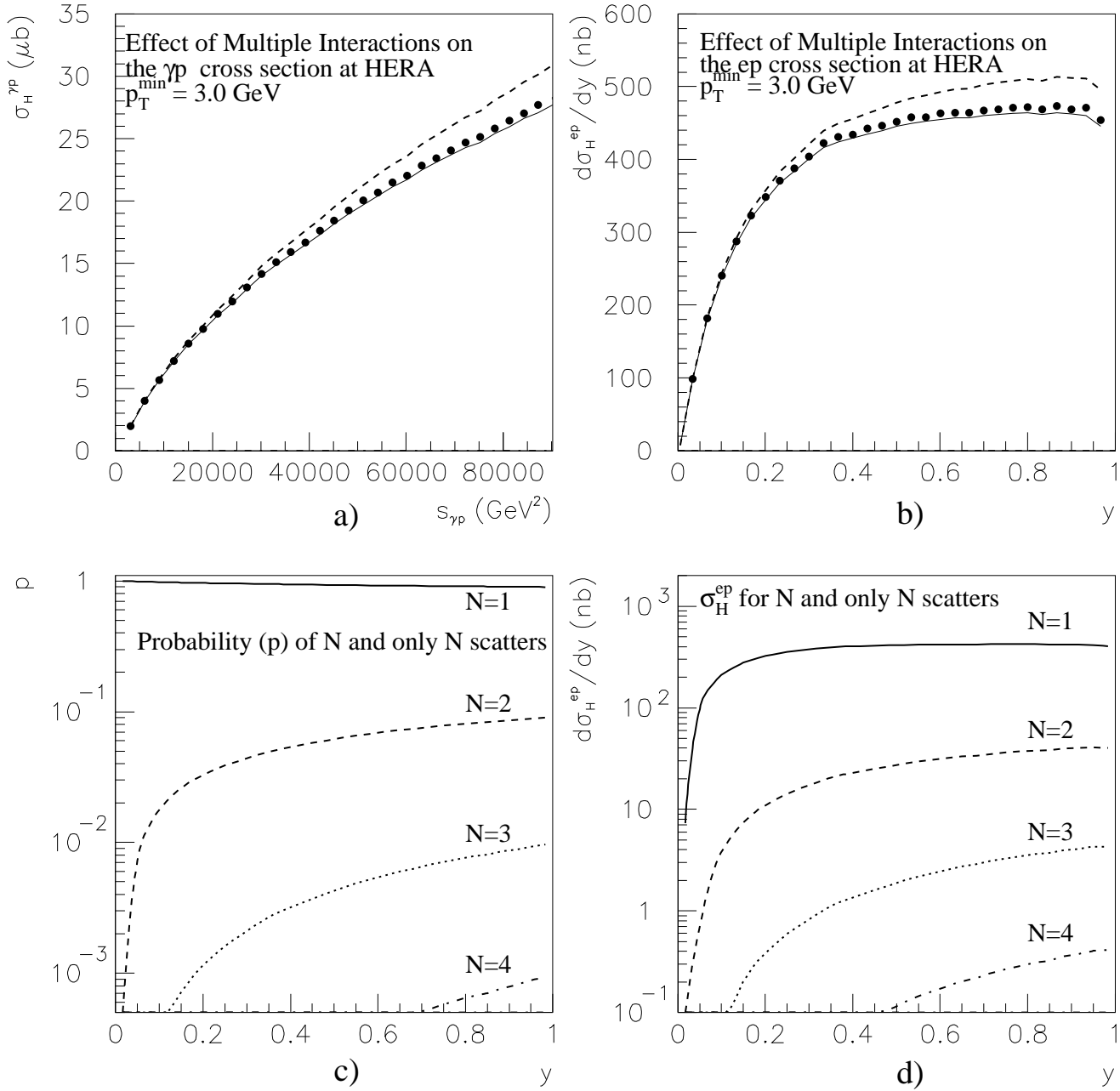}
\caption{\label{f:xsec} 
a) $\sigma_H$ as a function of $\sgp$, the squared 
c.m.\ energy of the photon-proton system.
b) $d\sigma^{ep}_H/dy$ as a function of $y$. In a) and b) the dashed lines 
show the result with no multiple interactions, the solid lines show the 
result of including multiple interactions and the points indicate the cross 
sections actually generated by the Monte Carlo program (see text).
In c) the probability of $N$ and only $N$ scatters as a function 
of $y$ is shown. In d) the cross section for $N$ and only $N$ scatters
as a function of $y$ is shown.}
\end{figure}

The effect of eikonalization  on the total $\gamma$-$p$ 
cross section is shown in figure~\ref{f:xsec}a. In figure~\ref{f:xsec}b 
this is translated into the effect on the differential $ep$ cross section
$d\sigma_{ep}/dy$ where $\sgp = 4 y E_e E_p$ ($E_e$ is the incoming
electron energy and $E_p$ the incoming proton energy).
The $\ptmin$ is taken to be 3~GeV and $\PRES = 1/300$. The parton 
distribution sets used are the GS2 set~\cite{GS} for the photon and the
GRV~\cite{GRV} set for the proton. Clearly (for this choice of parameters) 
the effect of multiple scattering is to significantly lower the cross 
section, with significance increasing with increasing $\sgp$ (and therefore 
$y$). This is reflected by a significant probability for more than
one hard interaction in a $\gamma p$ event, as shown in figure~\ref{f:xsec}c,
where it can be seen that, depending upon $y$, as many as 10-15\% of $\gamma p$
events contain more than one parton-parton scattering. This is expressed 
as a differential cross section for $N$ and only $N$ scatters in 
figure~\ref{f:xsec}d. Note that the hard cross section for $\surd\sgp 
\approx 200$~GeV represents around 20\% of the total 
$\gamma$-$p$ cross section \cite{SIGTOT}.

\subsection{Monte Carlo Implementation}

The cross sections illustrated in figure~\ref{f:xsec} indicate that 
the effects of multiple interactions might be observable at HERA\@. 
However, the experiments do not measure jet cross sections as low as 
$\ETJ \approx 3$~GeV. The lowest $\ETJ$ cut used on measured cross sections 
so far has been 6~GeV. The smearing of jet energies due to parton showering 
and hadronization effects mean that a $\ptmin$ cut of around 
2.5~GeV is often used by the experiments in Monte Carlo simulations to 
obtain a good description of the data at higher $\ETJ$ cuts.
The pile-up of more than one hard scatter in an event can be expected
to increase these migrations, and in order to study the effect of multiple
interactions on real jets rather than partons, it is necessary to include 
realistic hadronization and parton showering models in conjunction with
the eikonal model described above. To this end, the HERWIG Monte Carlo program
was adapted to allow the generation of multiple hard subprocesses in a single
hadron-hadron, photon-hadron or photon-photon event.

In the default version of HERWIG, event generation begins with the hard
process, QCD dijet production in our case.  This is generated according
to the lowest order cross section using whichever parton distributions
have been selected by the user.  Associated with the hard process is a
hard scale, $Q$ ($\sim\ETJ$ for dijet production), and the cross section
for additional partons to be radiated in the same event is enhanced by
powers of $\log Q/Q_0$ at each order in $\alpha_s,$ where $Q_0$ is a
lower cutoff defining the resolvability of partons.  In order to fully
describe the final state, $Q_0$ is required to be around the typical
hadronic scale, $Q_0\sim1$~GeV, and the logarithmic terms are large.
Therefore such terms must be summed to all orders in $\alpha_s$.  This
summation can be implemented as a probabilistic evolution of the
external (both incoming and outgoing) partons.  Any parton shower
algorithm that implements the DGLAP splitting functions \cite{DGLAP}
correctly resums
the single logarithmic terms associated with the collinear emission that
drives fragmentation function and parton distribution function
evolution, but to fully model the final state, single and double
logarithmic terms associated with soft emission must also be summed to
all orders.  This is done by {\em coherent parton shower
algorithms}~\cite{MW}, which also implement the DGLAP splitting
functions, but with specific choices for the definitions of evolution
scale and parton resolution criterion.  For final-state partons, the
algorithm evolves a single hard parton to many softer ones.  For
initial-state partons, the evolution is `backwards', from the hard
process to the incoming hadron or photon, evolving a single incoming
parton to another at larger~$x,$ together with many additional outgoing
partons.  The backward evolution algorithm~\cite{TorSjo} ensures that at
each stage of the evolution, the distribution of parton $x$ values
agrees with that of the input distribution function.  For the photon,
the inhomogeneous term in the evolution equation appears as an
additional vertex in the backward evolution, giving the possibility that
it will terminate `anomalously', with the incoming photon giving all of
its energy to the hard process and its associated parton shower.

Finally, after perturbative parton evolution, a non-perturbative model
of the transition from partons to hadrons is invoked.  In HERWIG, a
simple model that makes use of the preconfinement property of QCD is
used.  At leading order in the number of colours, the structure of a
parton shower is such that each parton ends up close, in both
momentum-space and real-space, to a parton that carries the opposite
colour quantum numbers.  The model therefore combines each such
colour-connected pair of partons to form a single colourless cluster.
This is assumed to decay directly to hadrons, with the choice of hadron
species controlled purely by phase-space arguments~\cite{Webb}.  This
model therefore has few adjustable parameters, which are already 
rather well constrained by data from e$^+$e$^-$ annihilation.
For processes with incoming partons, there is an additional freedom, of
how to model the break-up of the incoming hadron or photon after one
parton is taken from it.  In HERWIG, this is done by requiring that the
backward evolution results in a valence parton (u, d and s quarks and
antiquarks are all considered valence partons of a photon).  If the
perturbative evolution did not end anomalously, then additional emission 
is generated
below $Q_0,$ but still according to the perturbative distributions, to
obtain a valence parton.  Once the 4-momentum of this valence parton is
known, energy-momentum conservation is used to calculate the momentum of
the remaining hadron or photon remnant.  For photons, the flavour of the
remnant is simply that of the antiparticle of the valence parton, while for
protons, the remaining diquark is taken to be a single anticoloured
parton.  The cluster containing the remnant parton is then hadronized
like any other.  Clearly this step is unnecessary in events where the
backward evolution terminated anomalously, since there is no remnant
parton in that case.

When incorporating our multiple interactions model into the Monte Carlo
event generator several additional problems arise from the fact that the
model assumes that different scatterings are uncorrelated, while in the
event generator this is not possible.  The first is that of energy
conservation. In the analytical model, there is a non-zero (though
small) cross section for events in which the total energy participating
in hard scatters is greater than the energy of the incoming particles.
This is exacerbated by the backward evolution, since additional energy
is radiated away in the additional partons.  We impose global
energy-momentum conservation by adding the simplest possible correlation
between scatters: a $\Theta$-function in the cross section to produce
$m$ and only $m$ scatters requiring their total energy to be less than
the available energy.  This will reduce the amount of eikonalization of
the cross section relative to the analytical model.  When generating
events, this is implemented by calculating the total cross section and
probabilities for exactly $m$ scatters using the analytical model, and
generating hard scatters according to these probabilities.  Scatters in
which energy-momentum cannot be conserved are then vetoed (although the
event itself is kept, with all the other scatters).  At the end of event
generation, the number of vetoed scatters is used to make a revised
estimate of the total cross section.

In the analytical model, only events in which the photon is resolved are
eikonalized, while direct photon events are not.  However, events in
which the backward evolution terminated anomalously should also be
called direct in this context (they correspond to the small size
fluctuations of the photon) and should not be eikonalized.  Therefore,
if the backward evolution does terminate anomalously, no
further scatters are allowed in that event.  This is treated just like
the vetoing from energy conservation so also revises the final estimate
of the total cross section.  Since the separation into the anomalous and
resolved parts of the distribution function is model dependent, we
provide an option that enables multiple interactions in anomalous
events, so that one can gauge the relative importance of this feature of
the model.

These revised cross sections are shown as data points on
figure~\ref{f:xsec}a and b.

The simple model used in HERWIG for the hadron and photon remnants is no
longer appropriate when there is more than one scatter.  This is because
once the incoming colourless hadron has been replaced by an outgoing
coloured remnant by the first interaction, the same procedure cannot be
iterated for subsequent interactions.  Instead, we label the remnant as
a new kind of incoming hadron for the other interactions, which has
identical properties to the original hadron, except that gluons are
labelled as its valence partons.  Thus, if the backward evolution does
not result in a gluon, an additional emission is forced to produce one.
The outgoing coloured remnant from this interaction is the same as that
from the first, but with reduced momentum.  This procedure can then be
iterated as many times as required for all subsequent interactions.  By
thus modifying only the part of the model that connects the hard
processes to the incoming hadron, and not the backward evolution itself,
we ensure the predictivity of the model, using the parameters fixed to
other reactions.

\section{Effect on the Hadronic Final State}
\label{sec:sm}

The jet properties of the final state are of great importance in
understanding the underlying QCD processes, and the possible influence
of multiple
scattering needs to be carefully examined. For example, jet rates in 
photoproduction may well lead to important information regarding the 
gluon content of the photon and the proton~\cite{JEFF1,others}. 
In this section we examine the general effects of multiple scattering on the 
properties of the final state measurable at HERA\@. In order to do this, we 
take a default set of parameters chosen where appropriate to reflect 
published data from the HERA experiments and the best available theoretical 
estimates. These choices are as follows:
\begin{itemize}
\item $\PRES = 1/300$: This is motivated by assuming $\rho$-dominance. 
\item $\ptmin = 3.0$~GeV: This is consistent with the values typically used 
by the experiments to describe their data at $\ETJ \ge 6$~GeV.
\item The GRV parton distribution set is used for the proton~\cite{GRV}.
This set is the result of a global analysis including the HERA measurements of
the proton structure function $F_2$~\cite{DIS}.
\item The GS2 parton distribution set is used for the photon~\cite{GS}. 
The gluon distribution in the photon is poorly constrained over the region
studied (the desire to constrain the gluon is a major motivation for 
measuring high-$E_T$ photoproduction cross sections at HERA) and there is 
little reason to favour any one out of several available parameterizations.
\item A cut on the $\gamma p$ CM energy was made, 
114~GeV $\le \surd \sgp \le 295$~GeV, similar to those usually made by the 
experiments. 
\end{itemize}
These choices define what we call our `default' model. In the 
subsequent sections the sensitivity to various choices made here will be 
examined and discussed. Events were generated both with and without multiple 
interactions. 
All other parameters of the Monte Carlo model were left at their default
values, which are tuned to e$^+$e$^-$ data and give a good description
of the direct component of photoproduction at HERA~\cite{DIJETS}.

The initial expectation is that each pair of jets will be produced 
back-to-back azimuthally with balancing transverse momenta.
For two-to-two parton scattering in leading order QCD energy and 
momentum conservation gives

\begin{eqnarray}
\xplo = \frac{ \sum_{{\rm partons}}(E + p_z)_{{\rm parton}}} {2E_p}, \;
\xglo = \frac{ \sum_{{\rm partons}}(E - p_z)_{{\rm parton}}} {2yE_e}.
\end{eqnarray}

\noindent The proton direction defines the positive $z$-axis,
$E_e$ and $E_p$ are the initial lepton and proton energies and the sum is
over the two final state partons. For direct photon events, $\xglo = 1$. 
Since it is not possible to measure partons an observable has been 
defined by ZEUS~\cite{DIJETS} in terms of jets which is analogous to $\xglo$. 
This observable, called $\xgo$, is the fraction of the photon's momentum 
which emerges in the two highest $\ETJ$ jets. The explicit definition is,

\begin{equation}
\xgo = \frac{\sum_{{\rm jets}}E_T^{{\rm jet}}e^{-\eta^{{\rm jet}}}}{2yE_e},
\label{xgoeq}
\end{equation}

\noindent where now the sum runs over the two jets of highest $\ETJ$.
In the $\xgo$ distribution thus obtained, the LO direct and resolved processes 
populate different regions, with the direct processes concentrated at high 
values of $\xgo$. The peak arising from the direct contribution will not 
necessarily lie exactly at $\xgo = 1$ due to higher order effects and/or 
hadronization, but will still correspond to the kinematic region where most 
or all of the energy of the photon participated in the hardest scatter. 
Making a cut on $\xgo$ provides a workable definition of direct and
resolved events. Recall that beyond lowest order, such a separation is
ambiguous. We introduce a similar definition for the 
proton, i.e.

\begin{equation}
\xpo = \frac{\sum_{{\rm jets}}E_T^{{\rm jet}}e^{-\eta^{{\rm jet}}}}{2E_p}.
\label{xpoeq}
\end{equation}

In order to make our study as realistic as possible, 
we have performed jet finding on the hadronic final 
state using a cone algorithm. Unless stated explicitly otherwise, the 
cone radius used is $R = 1$ (as has been used by both HERA experiments so 
far), and jets have $E_T \ge 6$~GeV and pseudorapidity 
$\eta = -\ln(\tan\theta/2) \le 2$. These cuts place our jets well 
within the region observable by the HERA detectors. 

\begin{figure}
\vspace{5cm}
\setlength{\unitlength}{1mm}
\epsfysize=400pt
\epsfbox[50 150 450 550]{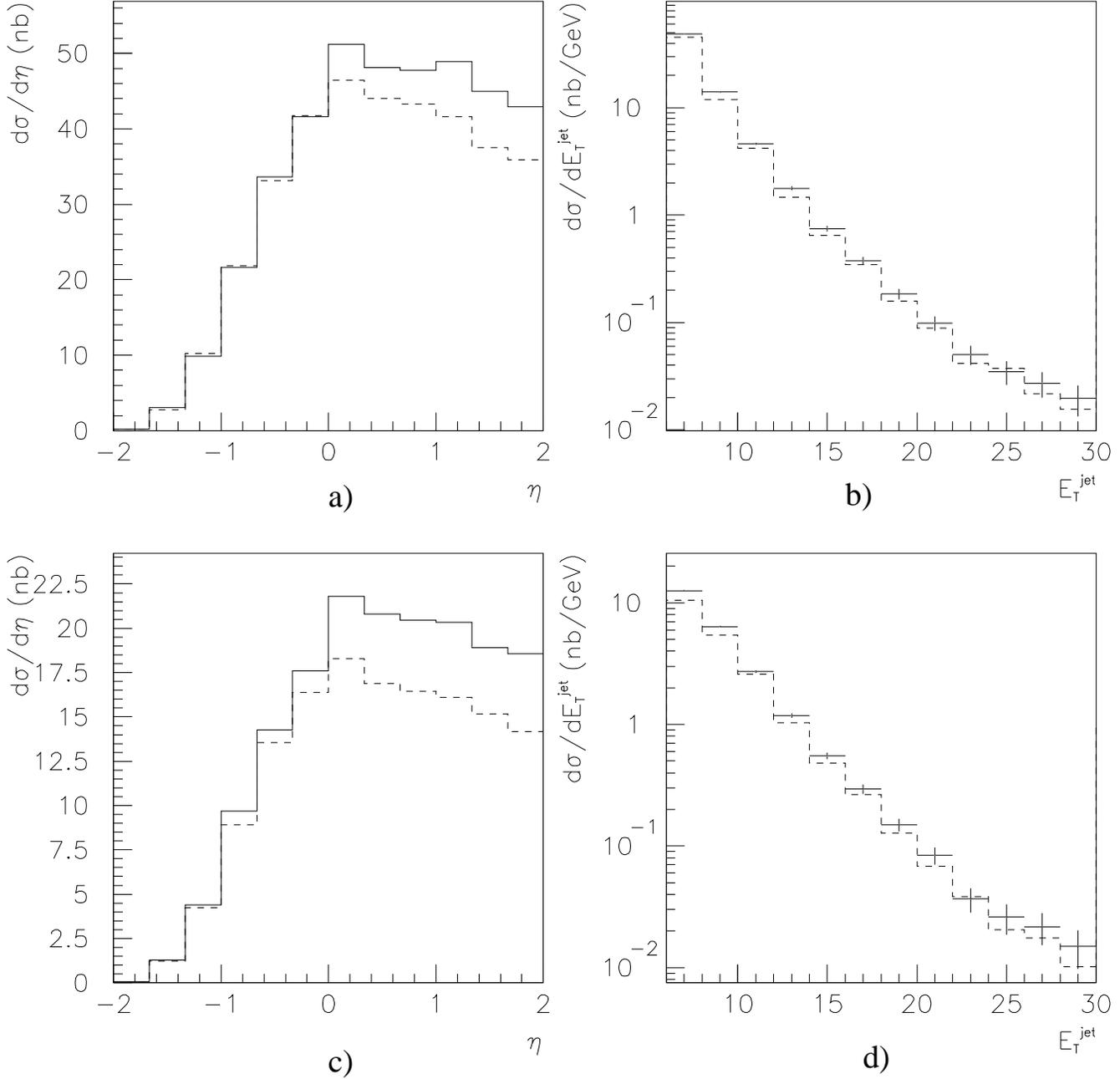}
\caption{\label{f:incdij}
Inclusive jet cross sections: a) $d\sigma/d\eta$ and b) $d\sigma/dE_T$.
Dijet cross sections: c) $d\sigma/d\eta$ and d) $d\sigma/dE_T$.
In all cases, $\ETJ > 6$~GeV.
The solid lines show the distributions when multiple scattering is included,
the dashed lines show the distributions when no multiple scattering is 
allowed.
In the $\ETJ$ plots, the  statistical errors are indicated on the
distribution for which multiple interactions are included. The errors on 
the `no multiple interactions' distribution are similar.}
\end{figure}

Multiple parton scattering is expected to affect jet rates in two ways.
Firstly, and most obviously, the average number of jets per event should 
be increased when partons from secondary hard scatters are of sufficiently 
high $p_T$ to give jets in their own right. In addition, there is also the 
possibility that extra hard scatters can influence the observation of jets 
even when no parton from the secondary scatters is of high enough $p_T$ to 
produce an observable jet. Lower $p_T$ secondary 
scatters produce extra transverse energy in the event which can contribute 
to the pedestal energy underneath other jets in the event. In this way, the 
inclusion of multiple scattering can be thought of as an attempt to extend 
perturbative QCD in order to calculate some part of the so called `soft 
underlying event', which is included in many simulations as a parameterized
extrapolation of existing data. The first of these effects will only increase 
multi-jet cross sections, whilst leaving the inclusive jet cross section 
unchanged. The second effect will increase both inclusive and multi-jet cross 
sections. In figure~\ref{f:incdij}, where inclusive jet 
(figure~\ref{f:incdij}a,b) and dijet 
(figure~\ref{f:incdij}c,d) 
cross sections are shown (as a function of jet rapidity and mean jet
rapidity respectively), it can be seen that the effect of multiple 
interactions is indeed more significant for the dijet cross sections than
the inclusive jet cross sections. 
In all cases the effect is more significant at lower $\ETJ$.
The direct photon contribution generated 
by HERWIG is included.

In table~\ref{t:rates} the calculated rates for one-, two-, three- and 
four-jet events in the `default model' are presented. The cross section for 
$m$ jets is defined here as the production cross section of 
events containing $m$ and only $m$ jets. There is an 
enhancement in all jet rates above the single jet rate and the significance 
rises with increasing jet multiplicity. Single jet events arise primarily
from configurations where the second jet is forward of the $\eta$ cut. 
Without considering the extra
transverse energy generated by multiple scattering, we would expect the
sum of the jet cross sections obtained including multiple scattering to be
lower than the corresponding sum obtained without multiple scattering
(and the latter should be concentrated more at lower jet rates), by a factor
equal to the mean multiplicity of jet pairs. This effect is mostly washed 
out by the large effect of the `underlying event' generated by the secondary 
scatters.  In this table only the contribution
from LO resolved photon processes is included. 
Since the parton shower model without multiple scattering only generates
multi-jet cross sections as a leading logarithmic dressing of the dijet
cross section, it is only accurate in the region in which two jets are
significantly harder than all the others.  Therefore the predicted rates
should only be taken as a rough indication.  Nevertheless, since the
logarithmically enhanced terms are included, the neglected terms are
genuinely suppressed by powers of $\alpha_s$, without additional enhancement,
whereas the eikonal expansion is an expansion (summed to all orders) in the 
large parameter $\alpha_s N^2$ (where $N^2$ is the product of the number 
densities).  We therefore expect the enhancement of higher jet rates to 
remain significant.

\begin{table}
\centering
\begin{tabular}{|l|c|c|}                \hline
             & \multicolumn{2}{|c|}{\bf Cross section (nb) }    \\ \hline
             & No multiple scatters & With multiple scatters    \\ \hline
  One jet    &   $90.8  \pm 0.6$      & $89.9  \pm 0.8$             \\ \hline  
  Two jet    &   $17.0  \pm 0.3$      & $19.5  \pm 0.4$             \\ \hline  
  Three jet  &   $0.73  \pm 0.06$     & $1.15  \pm 0.09$            \\ \hline  
  Four jet   &   $0.02  \pm 0.01$     & $0.05  \pm 0.02$            \\ \hline  
\end{tabular}
\caption{\label{t:rates} Resolved photon cross sections for multijet events. 
Jets have $\ETJ > 6$ GeV, $\eta < 2$, $R = 1$ and the proton and photon parton 
distribution sets used are GRV and GS2 respectively.
The errors shown are statistical.}
\end{table}

These results imply that, even with these rather strict cuts 
on the jets, around 500 four-jet events 
(where the jets are easily observable in the detector)
generated from multiple hard
interactions should be present in the data already taken by each of 
the HERA detectors (around 10~pb$^{-1}$ by the end of 1995).

\begin{figure}
\setlength{\unitlength}{1mm}
\epsfysize=400pt
\epsfbox[50 400 450 800]{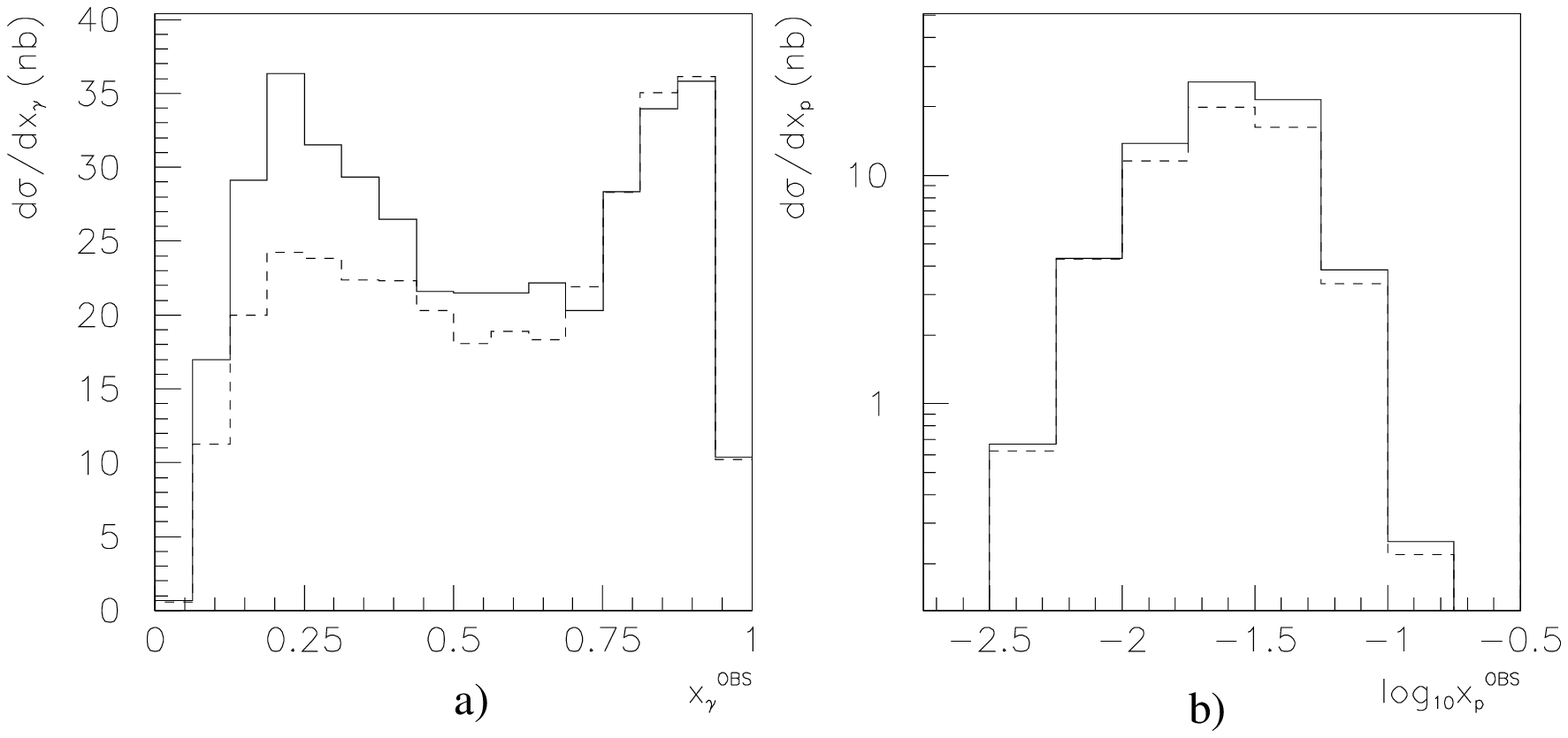}
\caption{\label{f:xgp}
a) $d\sigma/d\xgo\;$ b) $d\sigma/d\xpo$ 
The solid histograms show the distributions when multiple scattering is 
included, the broken histograms show the distributions when multiple 
scattering is neglected.} 
\end{figure}

Figure~\ref{f:xgp} shows the $\xgo$ and $\xpo$ distributions,
again including the direct contribution generated using HERWIG 
(which can be seen peaking at high values of $\xgo$).
The inclusion of multiple scattering raises the cross section in the
low $\xgo$ region without affecting the high $\xgo$ region, which is dominated 
by direct photoproduction. In the (uncorrected) $\xgo$ distributions
presented in refs.\cite{ZEUS2,DIJETS} there is an excess 
of data over the standard simulations in just this region. Multiple 
interactions may have a role to play in resolving this discrepancy.
The effect of multiple interactions can be seen to be largest for the
$x_p$ values in the middle of the available range. This is easy to understand
since the lower $x_p$ values are strongly correlated to large $x_\g$ values
(where direct photon interactions are dominant).

\begin{figure}
\vspace{-5cm}
\setlength{\unitlength}{1mm}
\epsfysize=400pt
\epsfbox[50 400 450 800]{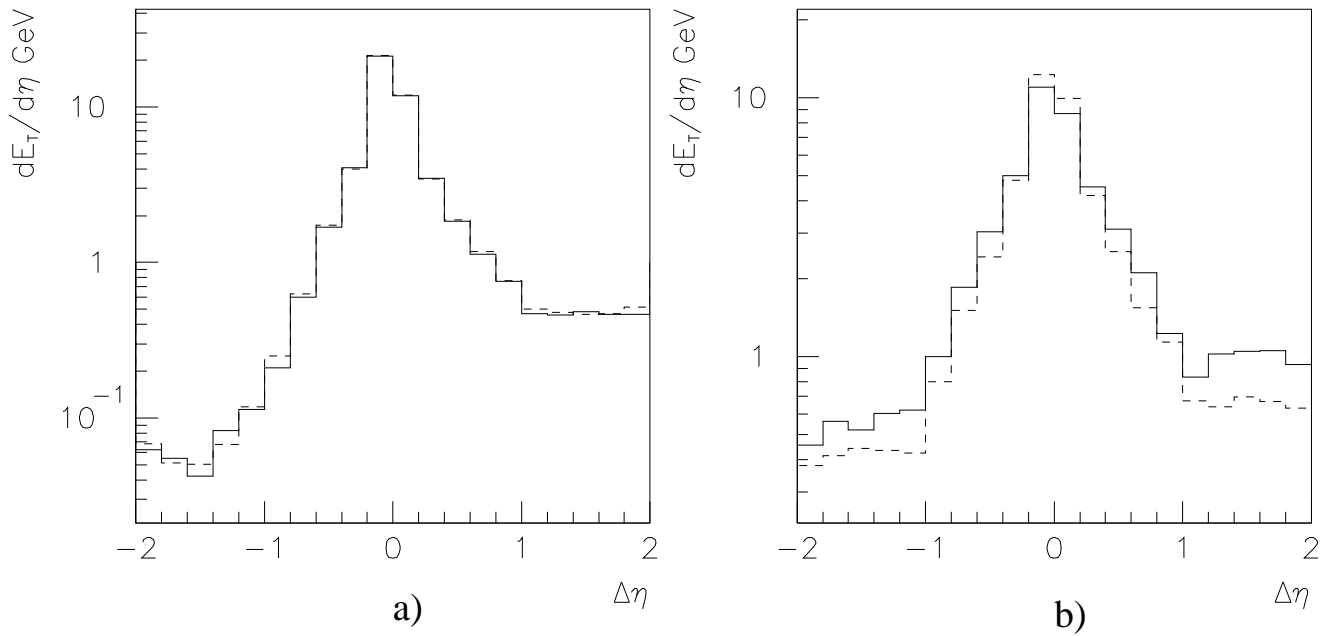}
\caption{\label{f:profs}
Transverse energy flow as a function of $\Delta\eta$ integrated over 
$|\Delta\phi| < 1.0$ 
for jets with $E_T > 6$~GeV in the range $-2.0 < \eta < 2.0$ 
and $\xgo \ge 0.75$ (a) and $\xgo < 0.75$ (b). 
The solid lines show the distributions when 
multiple scattering is included,
the dashed lines show the distributions when no multiple scattering is allowed.}
\end{figure}

Figure~\ref{f:profs} shows how multiple scattering affects the profiles 
of jets in the $\eta$-$\phi$ plane. In particular we look at jets in both 
the high and low $\xgo$ regions.

The effect of this model for multiple interactions on the jet profiles is 
significant but fairly small. 
In the high $\xgo$ region the effect is negligible. 
In the low $\xgo$ case, there is a slight broadening of the jet and an
increase in the pedestal of energy flow around the jet by around 300~MeV per 
unit of $\eta$ on the 
forward (proton) side of the jet and by around 100~MeV on the rear 
(photon) side.
In measured energy flows at HERA, an excess of energy flow in the forward 
region has been observed~\cite{ZEUSinc} and shown to occur principally at
low values of $\xgo$~\cite{DIJETS}. None of the jet profiles
presented by the HERA experiments are corrected for detector effects,
and so smearing of jet and particle energies makes direct comparison
impossible at present. One might take the view that the jet profile data
could be used to tune $\ptmin$. This is not a very sensible thing to do,
since (as we shall see) uncertainties in the photon gluon density can
easily be traded off against the value of $\ptmin$.

\section{Sensitivity to Parton Distributions}
\label{sec:pdfs}

\begin{figure}
\setlength{\unitlength}{1mm}
\epsfysize=400pt
\epsfbox[50 150 450 550]{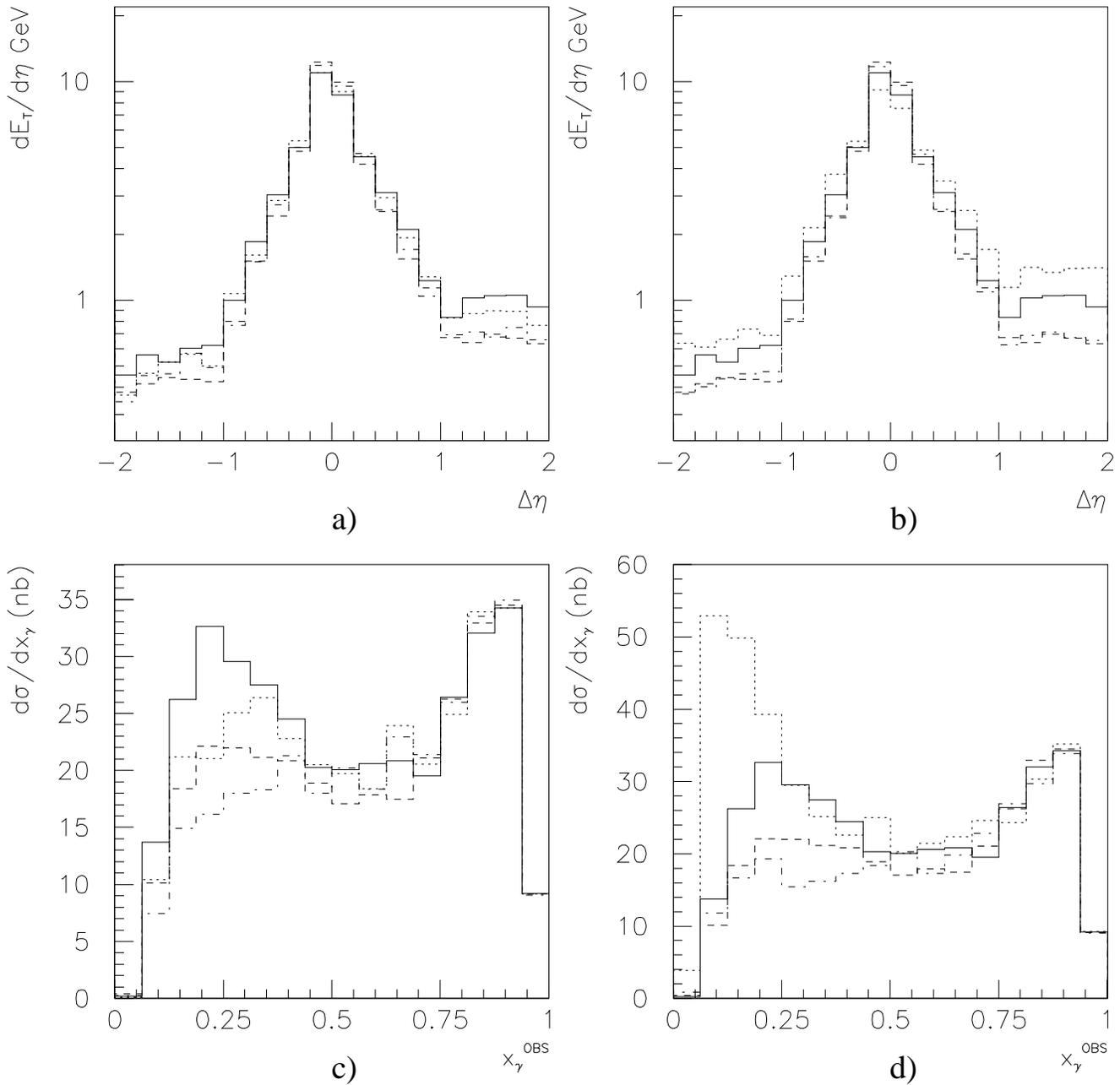}
\caption{\label{f:sfs} 
Effect of different photon parton distribution sets: 
a) and b) show transverse energy flow as a function of $\Delta\eta$ 
integrated over $|\Delta\phi| < 1.0$ for jets with $E_T > 6$~GeV in the 
range $-2.0 < \eta < 2.0$ and $\xgo \le 0.75$. 
The solid and dashed lines show the distributions obtained  using the GS2 
photon parton densities with and without multiple interaction respectively.
In a) the dotted and dash-dotted lines show the distributions obtained using 
the GRV photon parton densities with and without multiple interactions 
respectively, and in b) they show the same distributions obtained using the 
LAC1 photon parton densities. Figures c) and d) show the $\xgo$ cross section. 
The lines in c) have the same meaning as in a), and those in d) have the same 
meaning as in b). In all cases the proton parton distribution set is 
taken from GRV.}
\end{figure}

The fact that jet cross sections, and in particular dijet cross sections, 
are affected by the presence of multiple hard interactions means that 
these cross section may be misleading if used to distinguish between
different parton distributions in the proton and/or photon without giving
due consideration to the effects of multiple interactions.
In figure~\ref{f:sfs} we show the $\xgo$\ distribution and the
low-$\xgo$ jet profiles obtained using the GS2, LAC1 and GRV 
parton distribution sets for the photon.

In general, for this kinematic region, the effect of multiple scattering is 
greater than or similar to the changes in the cross sections produced by using 
different parton distributions. This is particularly notable in the jet
profiles,
which in the absence of multiple interactions are insensitive to the photon 
parton distribution, but which depend upon it significantly when
multiple interactions are allowed. The effect of multiple interactions is most 
pronounced if the LAC1 parton distribution set is used. 
This is due to the high density of low-$x_{\g}$ gluons present in the 
parameterization. In table~\ref{t:lac1} we show the 
rates of multijet events for the LAC1 photon parton distribution set,
where it can be seen that the multijet rates are massively enhanced by
multiple interactions.

\begin{table}
\centering
\begin{tabular}{|l|c|c|}                \hline
             & \multicolumn{2}{|c|}{\bf Cross section (nb) }    \\ \hline
             & No multiple scatters & With multiple scatters    \\ \hline
  One jet    &  $93.2  \pm 1.0$       &   $124.4  \pm 1.2 $         \\ \hline  
  Two jet    &  $15.0  \pm 0.4$       &   $28.3  \pm 0.6 $         \\ \hline  
  Three jet  &  $0.62  \pm 0.08$      &   $2.13  \pm 0.16$         \\ \hline  
  Four jet   &  $0.01  \pm 0.01$      &   $0.15  \pm 0.04$         \\ \hline  
\end{tabular}
\caption{\label{t:lac1} As Table~1, but using LAC1 parton
  distribution functions for the photon instead of GS2.}
\end{table}

\section{Variations on the Default Model}
\label{sec:vary}

\begin{figure}
\setlength{\unitlength}{1mm}
\epsfysize=400pt
\epsfbox[50 150 450 550]{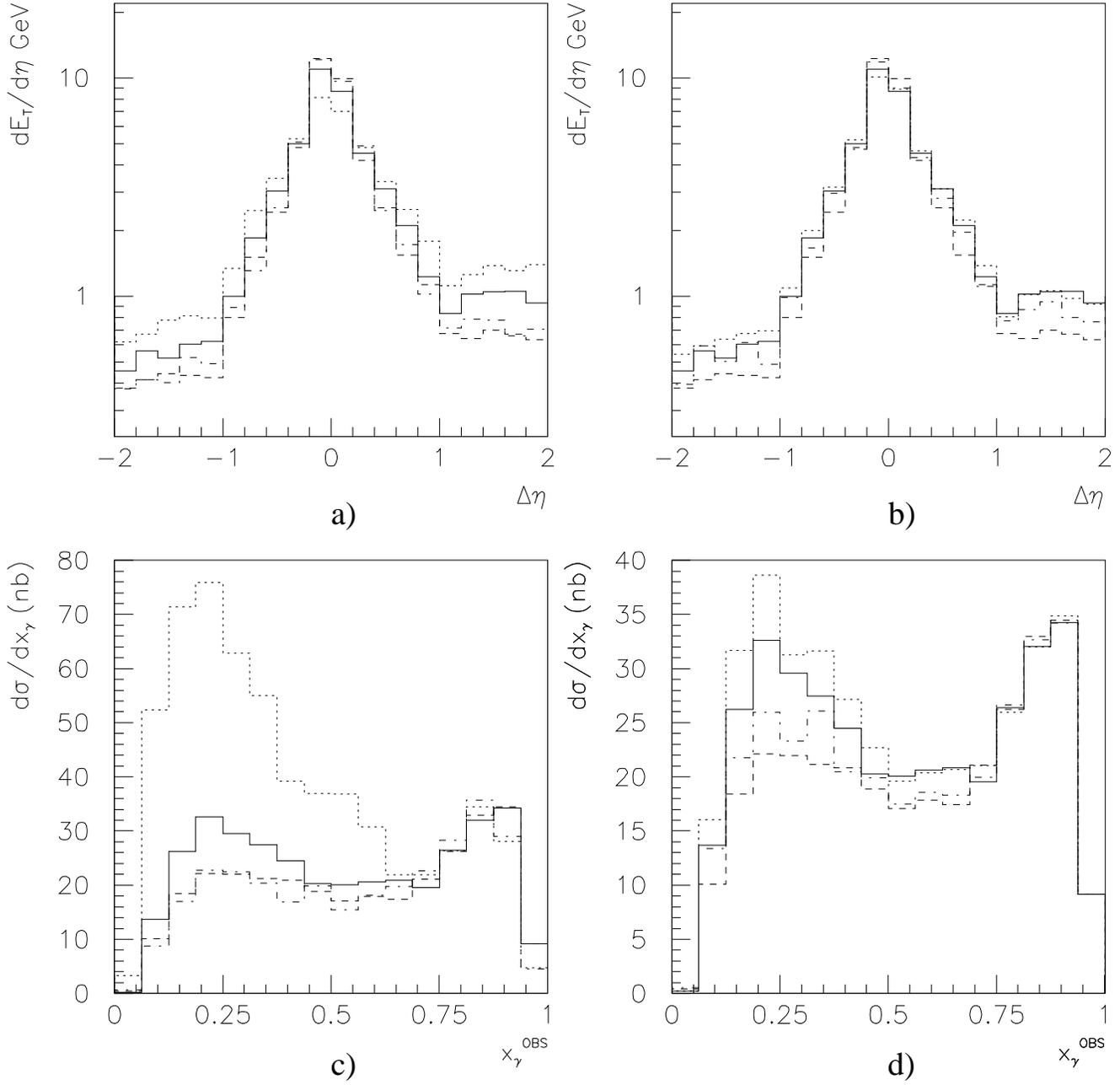}
\caption{\label{f:vary}
Sensitivity to model parameters: 
a) and b) show transverse energy flow as a function of $\Delta\eta$ 
integrated over $|\Delta\phi| < 1.0$ for jets with $E_T > 6$~GeV in the 
range $-2.0 < \eta < 2.0$ and $\xgo \le 0.75$. 
In all cases, the solid and dashed lines show the distributions obtained with 
and without multiple interactions respectively in the default model).
In a) the dotted and dash-dotted lines show the distributions obtained with 
and without multiple interactions respectively, when $\ptmin$ is lowered from 
3~GeV to 2~GeV. In b) the dotted line shows the distribution obtained with 
multiple interactions when the photon radius $\nu^2$ is changed to 
1.5~GeV$^2$ and the dash-dotted lines show that obtained with $\PRES = 1/150$. 
Figures c) and d) show the $\xgo$ cross section. 
The lines in c) have the same meaning as in a), and those in d) have the same 
meaning as in b).}
\end{figure}

Despite the fact that it is a reasonably simple and well defined procedure, 
there are of course several uncertainties and free parameters in the model. 
These include the values of $\ptmin$, $\PRES$, and  $\nu^2$. 
The parameter which has the strongest effect of the cross sections is the
value of $\ptmin$. Lowering $\ptmin$ to 2~GeV dramatically increases the 
effect of multiple interactions, as shown by the effect on the jet profiles and
and $\xgo$ distribution in figures~\ref{f:vary}a and c. The energy flow around
the jet axis and the cross section at low $\xgo$ increase 
when $\ptmin$ is lowered. This occurs whether or not multiple 
interactions are included, but in the presence of multiple 
interactions the effect is much greater.

The value of $\PRES$ used in our `default model' is 
around the lowest sensible value. Increasing it from $1/300$ to $1/150$ 
decreases the effect of multiple scattering since we now 
correspondingly reduce the number density of resolved partons.
Since $\PRES$ is the probability that a photon is resolved, its value is
closely related to the photon's parton distribution functions.  Indeed,
if one fixed the parton densities in a resolved photon and only varied
$\PRES,$ there would be no variation in the effect of multiple
interactions.  However, the biggest differences between current sets is
in the distribution functions within a resolved photon, rather than in
the value of $\PRES$ used.  Therefore most of the differences shown in
section~\ref{sec:pdfs} would remain even if $\PRES$ was adjusted to the
values in the distribution functions.

Changing the value of $\nu^2$ from 0.47~GeV$^2$ to 1.5~GeV$^2$ makes the
photon more compact, i.e.\ it increases 
the resolved-$\g$ number densities at small impact parameters whilst 
decreasing them at large impact parameters. 
Since peripheral $\g p$ collisions are rarer, the net result is to increases 
the effect of multiple interactions for
each resolved-$\gamma$--$p$ interaction but to make the cross section for
such reactions smaller. The effects of these changes on the 
jet profiles and $\xgo$ distribution are shown in figures~\ref{f:vary}b and d.

\section{Comparison with Data}

\begin{figure}
\setlength{\unitlength}{1mm}
\epsfysize=400pt
\epsfbox[50 150 450 550]{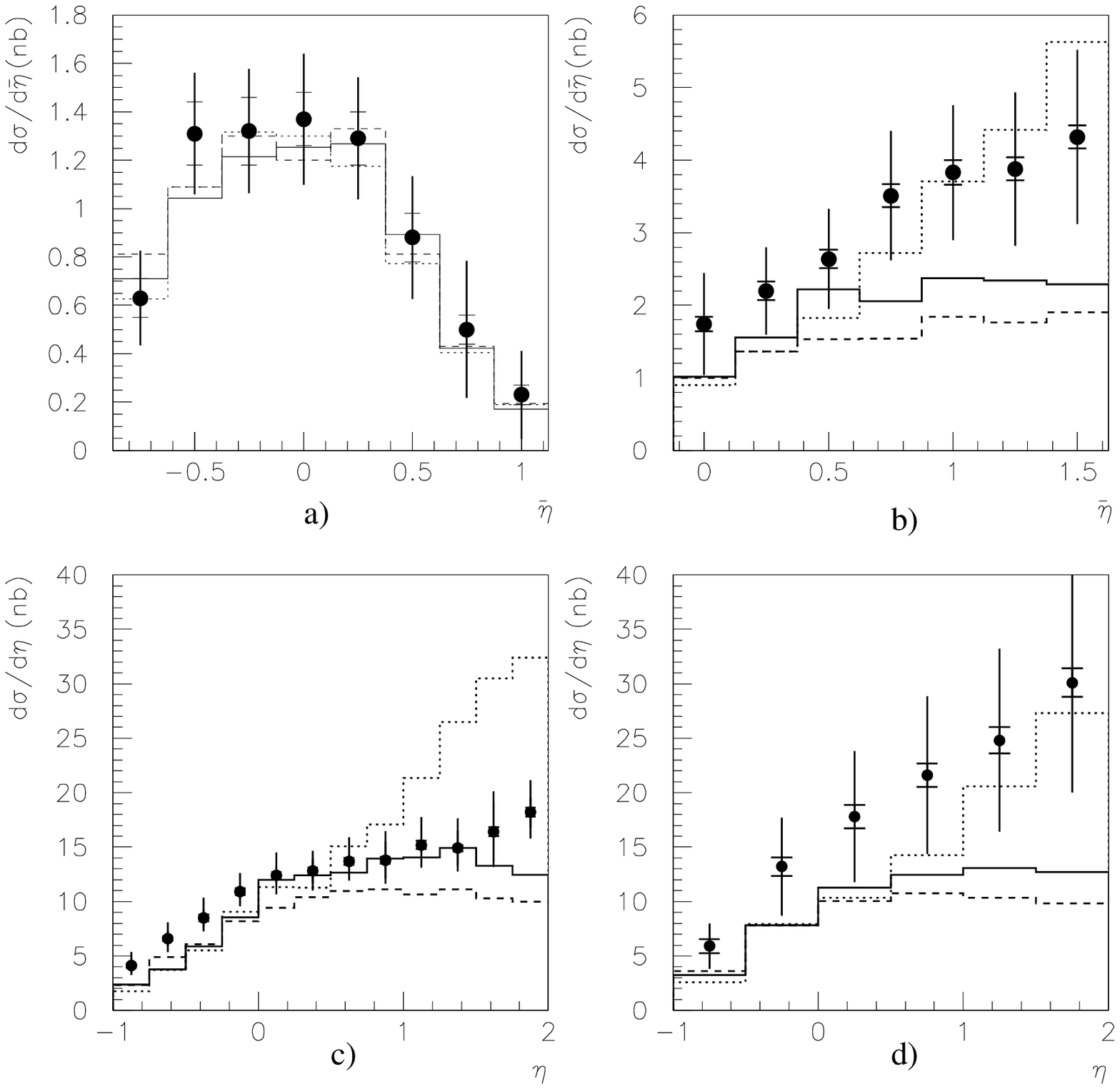}
\caption[]{\label{f:data1}
ZEUS data compared with HERWIG.
a) and b) show the differential dijet cross sections for direct and resolved
photon events respectively (where the separation is defined by a cut on 
$\xgo$) by the ZEUS collaboration~\cite{DIJETS}. 
c) and d) shows the differential inclusive jet cross section as 
a function of $\ETAJ$ as measured by the 
ZEUS~\cite{ZEUSinc} and  H1~\cite{H1inc2} collaborations respectively.
The inner error bars indicate the statistical uncertainty, and the outer 
one the statistical and systematic errors added in quadrature.
In all cases the solid histogram shows the calculation of the default multiple 
interaction model, the dotted line shows the calculation using the
default but using LAC1 instead of GS2 for the parton distribution set,
and the dashed line shows the result of HERWIG with no multiple interactions.}
\end{figure}

Both ZEUS~\cite{ZEUSinc} and H1~\cite{H1inc2} have published inclusive 
jet cross sections for photoproduction at HERA, and ZEUS has also published 
dijet cross sections~\cite{DIJETS}. Our default multiple
interactions model is compared to a selection of this available data in 
figure~\ref{f:data1}.

The dijet cross sections are measured as a function of the average
$\eta$ of the two jets ($\bar{\eta}$) for direct and resolved 
photoproduction (defined by a cut at $\xgo = 0.75$) for $\ETJ > 6$~GeV, 
$P^2 < 4$~GeV$^2, 0.2 < y < 0.8$ 
and $\DETA < 0.5$ ($P^2$ is the photon virtuality).
There is a correlated uncertainty due to the calorimeter energy 
scale of around 20\% in the direct and 25\% in the resolved data, 
which here we have added in quadrature to the other systematic errors.
The agreement with the direct dijet cross section is good in all cases,
confirming that the separation based upon $\xgo$ removes to a large extent
any sensitivity of this cross section to the photon structure (and hence
multiple interactions). In the case of the resolved dijet cross section, the 
calculations are in general too low, 
as was also the case for the analytic LO QCD calculations presented in 
\cite{DIJETS}. Multiple interactions raise the calculated
cross section towards the data and, given the large systematic errors on the
data (in particular the overall normalization uncertainty) perhaps the 
disagreement is not so significant yet. 
A major source of the systematic errors in the data is the
discrepancy in the jet profiles between data and Monte Carlo models, it is to
be hoped that improvements in the simulations (of which the inclusion
of multiple interactions is an example) will contribute to the reduction
of these errors in future measurements.

The inclusive single jet measurements of H1 and ZEUS are made over different 
$y$ ranges ($0.2 < y < 0.85$ for ZEUS and $0.25 < y < 0.7$ for H1), different
$P^2$ ranges ($P^2 < 4$~GeV$^2$ for ZEUS and $P^2 < 0.01$ for H1) and
different $\ETJ$ ranges ($\ETJ > 8$GeV for ZEUS and $\ETJ > 7$GeV for H1).
There is a correlated uncertainty due to the calorimeter energy 
scale of around 20\% in the ZEUS inclusive jet data, and a 
25\% overall normalization uncertainty in the H1 data.
In both cases we have here added these in quadrature to the other 
systematic errors.
In addition it is worth noting here
that since our model uses only the LO matrix elements plus parton showers,
there is an expected uncertainty in the calculations due principally to the
choice of scale. In the calculations available for these processes, the NLO 
corrections have been as high as $40\%$ \cite{Kramer} although, as
mentioned earlier, the parton shower would be expected to reproduce at
least part of this correction.
Since the ZEUS data are collected up to larger values of $y$, multiple
interactions are expected to have a larger effect and this is seen by
comparing the solid and dotted curves in figure~8c with the corresponding
curves in figure~8d. The broken lines, corresponding to the HERWIG predictions
without multiple scattering are very similar (the effect of increasing
the $P^2$ range in the ZEUS data is compensated by the lower $\ETJ$ cut
of H1). Given the uncertainties mentioned above, we should be cautious in
drawing any conclusions from these data. However, there is a suggestion
- at the $(1 \sim 2)\sigma$ level - that the HERA data may be incompatible,
in that H1 inclusive jets agree better with LAC1 distribution (or very 
significant multiple interaction effects) whilst the ZEUS inclusive jet 
data lie closer to the result of our default multiple interaction model. 

Bearing in mind the above discussions, there are signs that including 
multiple interactions can bring the model into better agreement with existing 
data in those regions where discrepancies exist
(i.e.\ the forward region in both dijet and inclusive jet production), 
whilst having a small effect on those regions
(i.e.\ direct dijet, low $\ETAJ$ inclusive) where the agreement 
is already good.

\section{Conclusions}

We have performed a detailed study of the unitarization corrections which are
expected to appear in high energy $\gamma p$ interactions. We used an
eikonal approach to model their effects which, in the perturbative domain,
manifest themselves through the appearance of multiple parton scattering. 
By performing a Monte Carlo simulation of our model we have been able to
make a study of the detailed properties of the hadronic final state which
enables us to make use of the vast amount of HERA data on photoproduction.
We find that, for partons produced with $\pT > 3$ GeV, over 4\% of events can 
be expected to contain more than one hard process at typical HERA energies. 

For reasonable experimental cuts the effect of multiple parton interactions 
on measured jet cross sections could be significant. The size of
the effect is expected to depend strongly upon the parton distributions in the
photon and can be as high as 100\% in some regions for dijet 
cross sections already measured at HERA\@. 
The effect is even more significant for higher jet rates, 
leading to an {\it overall\/}
enhancement of a factor of up to around five in the four jet rate 
(with all four jets contained in a detector).  These events 
should provide a means of unambiguously discovering or ruling out some 
multiple interaction models in the near future, although untangling
a clear signature in a realistic multi-hadron final state 
remains something of a challenge.

The model considered here also leads to the generation of a \lq semi-hard'
underlying event. This affects the energy flow forward of 
the jet direction, as do other models which are already being 
used to to describe HERA data \cite{PYTHIA,Engel}, where the energy flow in 
the forward (proton) direction is poorly reproduced in simulations which do 
not include multiple interactions. 

Collectively, the effects of multiple interactions may well make the
extraction of photon parton distribution functions (especially the gluon) 
rather difficult.

The Monte Carlo program used here runs in conjunction with HERWIG version 5.8 
and is available from the authors.  Further information can be obtained
from the world wide web page
\verb+http://surya11.cern.ch/users/seymour/herwig/+

\section{Acknowledgements}
It is a pleasure to thank DESY (where much of this work was performed)
for their hospitality. For useful discussions we thank Greg Feild, 
Lutz Feld, John Storrow and Rod Walker.

\end{document}